# A Time Efficient Indexing Scheme for Complex Spatiotemporal Retrieval

Lagogiannis G.[1], Lorentzos N.[1], Sioutas S.[3], Theodoridis E.[2]

[1] *Science Dep., Agricultural University of Athens, Iera Odos 75, 11855 Athens, Greece*

[2] *Computer Engineering and Informatics Dept. University of Patras, Greece*

[3] *Dep. Informatics, Ionian University, Corfu, Greece*

e-mails : {lagogian, lorentzos}@aua.gr, sioutas@ionio.gr, theodori@ceid.upatras.gr

**Abstract**

*The paper is concerned with the time efficient processing of spatiotemporal predicates, i.e. spatial predicates associated with an exact temporal constraint. A set of such predicates forms a buffer query or a Spatio-temporal Pattern (STP) Query with time. In the more general case of an STP query, the temporal dimension is introduced via the relative order of the spatial predicates (STP queries with order). Therefore, the efficient processing of a spatiotemporal predicate is crucial for the efficient implementation of more complex queries of practical interest. We propose an extension of a known approach, suitable for processing spatial predicates, which has been used for the efficient manipulation of STP queries with order. The extended method is supported by efficient indexing structures. We also provide experimental results that show the efficiency of the technique.*

## 1. Introduction

The efficient handling of spatiotemporal data is an increasing demand of modern DBMSs, motivated by location based services (e.g. GIS applications) and telecommunications (cellular networks). Such applications soon expanded in areas such as robotics, medical imaging, multimedia applications, etc [5]. Spatial attributes can be viewed as 0D, 1D, 2D or 3D positions. Temporal attributes capture the temporal existence of entities and, in the general case, they can be represented as time points or time intervals. The most typical form of spatio-temporal data is that of trajectories. A spatio-temporal predicate is a pair (S,T), where S represents a spatial constraint and T represents a temporal constraint, which can be either a time-instant t or a time interval Δt. A query of the form

$$Q_1 = \{(S_1,T_1),(S_2,T_2), \ldots, (S_N,T_N),\} \quad (1)$$

is referred to as *STP query with time*.
Spatio-Temporal Pattern (STP) queries [6] depend on the efficient manipulation of such predicates.

This paper is concerned with the efficient processing of such queries by using an appropriate indexing.

We say that a spatio-temporal predicate (S, T) is satisfied by the trajectory of an object if the object lies in S at some time within the specified temporal constraint T. Moreover, we say that the trajectory satisfies a query Q if it satisfies all the spatio-temporal predicates of Q.

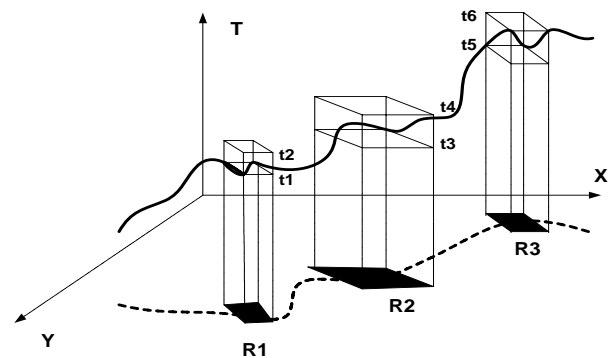

**Figure 1. An example query.**

As an example, the query depicted in Figure 1 is

$$Q=\{(R_1, [t_1,t_2]),(R_2, [t_3,t_4]), (R_3, [t_5,t_6])\}.$$

and it is satisfied by every trajectory that crosses the regions $R_1$, $R_2$ and $R_3$ at some time between $[t_1,t_2]$, $[t_3,t_4]$ and $[t_5,t_6]$, respectively.

The solution we propose in this paper is based on another solution [6], suitable for the efficient evaluation of *STP queries with order*. In these queries, the spatial predicates are not associated with temporal constraints. Instead, the dimension of time is inserted into the query via the order of the spatial predicates. Such an example query is

$$Q_2 = \{(S_1),(S_2), \ldots, (S_N)\} \quad (2)$$

The output of this query consists of all the objects that visited the areas $S_1, S_2, \ldots, S_r$ in this order.

## 2. Related work

The problem of indexing and querying spatio-temporal data lately has gained much attention. Güting et al [4] propose a data model and a query language for handling and expressing complex spatio-temporal queries. Several trajectory-indexing methods have also been proposed for the handling of spatial predicate queries (see [10] for a survey). Theodoridis et al. [12] study the issues that arise in spatio-temporal index structures. Chakka et al [3] propose a two-level method that decouples the indexing of the spatial and the temporal dimensions of the datasets. Pfoser et al [11] propose two access methods, the STR-tree and the TB-tree. The former is based on the classical $R^*$-tree [2] whereas the latter is an R-tree hybrid structure that preserves trajectories.

A different approach for the handling of STP queries with order has been introduced in [6]. The idea considers a grid on a 2-dimensional space. Each cell of the grid is associated with a list. Assuming in particular that an object $O_l$ enters cell $C_m$ at time $t_k$, the pair ($O_l$, $t_k$) is inserted into a list associated with cell $C_m$. Each such list is ordered by object Id. Given also that an object may enter a cell more than once, all the times of entrance of an object in the same cell are ordered by time. A simple example of the approach is depicted in Figure 2, and explanations are as follows:

The arrows on a trajectory show the direction of movement of an object. For simplicity, it is assumed that an object remains in a cell for a single time-instant. Each bullet on the trajectory of an object denotes the time instant at which the object sends a message. Hence, the Figure shows that object $O_1$ entered cell $C_2$ at time 3. Similarly, object $O_2$ entered $C_2$ at time 8 and also at time 10. Each element of a list is called *record*.

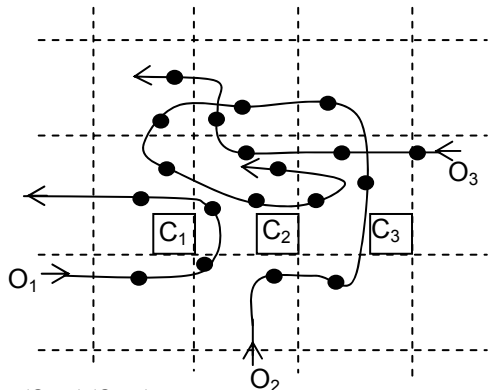

$C_1$: ($O_1$, 4) ($O_2$, 7)
$C_2$: ($O_1$, 3) ($O_2$, 8) ($O_2$, 10) ($O_3$, 3)
$C_3$: ($O_2$, 3) ($O_2$, 9) ($O_3$, 2)

**Figure 2. A partitioning of space into cells and representation of lists**

In an STP query with order (expression (2)), all the predicates are range constraints, and they are evaluated concurrently, by merging the lists associated with these predicates. The answers are retrieved in sorted object identifier order. In the remainder of the paper, we refer to this approach as *the list solution.*

For each spatial or spatio-temporal predicate $P_i$, let $F(P_i)$ be the set of object Ids which satisfy $P_i$. Let also B be the number of records that fit in a block of secondary memory. Finally let N be the number of predicates of the query. Assuming that the predicates are processed sequentially, an obvious upper bound on the number of I/Os for the evaluation of the query is

$$O(\max(|F(P_i)|, 1 \le i \le N)*N/B),$$

where $|F(P_i)|$ is the cardinality of $F(P_i)$. This is because every object belonging to $F(P_i)$ has to be examined in order to find out when it satisfies all the remainder predicates.

As is going to be shown later, the list solution achieves this bound for STP queries with order but it does not work efficiently for STP queries with time.

To efficiently manipulate spatio-temporal predicates, for STP queries with time, in this paper we propose the use of persistent indexing structures. To the best of our knowledge, persistent techniques have not been studied for this purpose.

In the remainder of this paper we make the following assumptions / simplifications.
1. Every spatial predicate matches a cell. Note that although this is rarely the case, it does not affect the efficiency of our solution. In addition, it allows to concentrate on the indexing structures.
2. An object does not enter a cell more than once during the time interval specified in the query. This is a realistic assumption, if the time predicate does not represent an extremely long time interval. For example, the percentage of vehicles entering a certain cell more than once, during a period of a few hours, is expected to be extremely low.

Assume now that an object $O_i$ enters cell $C_k$ at time instances 6, 10, and 15. It is then noted that, at time instance 8, it cannot be determined whether $O_i$ is still in $C_k$. This is because we do not know the time at which $O_i$ left $C_k$. To overcome this problem, if object $O_i$ leaves $C_k$ at time 7, then a record ($O_i$, 7) is stored in the list of $C_k$. Such records appear shaded in the remainder of the paper. Note that the knowledge of exit of an object from a cell is not needed for STP queries with order. Due to this, there is no need to maintain exit records in the lists of Figure 2.

$C_1$: ($O_1$,4) ($O_1$,5) ($O_2$,7) ($O_2$,8)
$C_2$: ($O_1$,3) ($O_1$,4) ($O_2$,8) ($O_2$,9) ($O_2$,10) ($O_2$,11) ($O_3$,3) ($O_3$,4)
$C_3$: ($O_2$,3) ($O_2$,4) ($O_2$,9) ($O_2$,10) ($O_3$,2) ($O_3$,3)

**Figure 3. The new lists of Figure 2**

By the introduction of exit records, the list of Figure 2 could have the form shown in Figure 3.

## 3. Depicting the inefficiency

When dealing with STP queries with order, a predicate has first to be chosen for evaluation. Let this predicate be $C_1$. It then suffices to use lists for the storage of the objects that are inside the cells. Indeed, all the nodes in the list contain objects that may belong to $F(C_1)$, therefore they all have to be retrieved.

Now assume that the query in discussion is:
$$Q=\{(C_2, 6\text{-}8), (C_1, 7), (C_3,9)\}$$
according to the lists of Figure 3. Since the predicates are spatio-temporal, Q is an STP query with time. Consequently, the objective of this query is to find all the objects that entered cell $C_2$ between the time instants 6 and 8, they were in cell $C_1$ at time 7, and they were in $C_3$ at time 9.

By using the list solution, we then have to examine all the objects contained in the list, i.e. $O_1$, $O_2$ and $O_3$, despite the fact that $F(C_2, 6\text{-}8) = \{O_2\}$. Obviously, such an examination is problematic if the number of objects inside each cell is large. Indeed, in such a case, the cell lists are too long, and a single I/O does not suffice to retrieve the entire list. In the worst case, one I/O per object will be needed, provided that pointers have been used, to point to the first occurrence of every object. It follows, therefore that, for an efficient processing of a query, in $O(\max(|F(P_i)|, 1 \leq i \leq N)*N/B)$ I/Os, a different approach has to be developed. Indeed, this is the objective of this paper. A new solution is proposed in Section 5, which enables the processing of a spatio-temporal predicate P by consuming $O(|F(P)|/B)$ I/Os and a satisfactory space consumption. Before this solution is described, another primitive solution is also presented in Section 4, which, however, suffers from enormous space consumption.

## 4. The primitive solution

In this approach, each cell $C_i$ is associated with two structures, S*tructure A* and S*tructure B.*

Structure A is a two-level structure: The upper level is an index for the object Ids. Each leaf (object Id) is associated with another index at the lower level, with the times at which this object entered cell C.

Structure B is also a two-level structure: The upper level is an index for time stamps. Each leaf (time instant) is associated with a list (lower level) containing the object Ids that were in cell $C_i$ at the given instant.

Now, let
$$Q=\{(C_1,T_1),(C_2,T_2),\ldots,(C_r,T_r)\}$$
be the query. Initially, the upper level of Structure B of cell $C_1$ is searched, in order to find the time-stamps, which satisfy the temporal predicate $T_1$. We then follow the corresponding list of the lower level and we store the object Ids into a set V. Next, for each distinct object $O_i$ in V, we have to check whether it satisfies all the remainder predicates.

Let therefore $(C_2, T_2)$ be the next predicate and assume that $T_2$ is time $t_2$. To check whether object $O_i$ satisfies this predicate, we make use of Structure A of $C_2$. The path of the upper level of Structure A leading to object $O_i$ and the leaf corresponding to $O_i$ is connected with a lower level indexing structure that contains the time instants at which $O_i$ was in $C_2$. Hence, we can find whether $O_i$ was in $C_2$ at time $t_2$.

The major advantage of this solution is its time efficiency, compared to the list solution. To make this clear, consider a cell in which a large number of objects have entered but, at each time instant, the number of objects in the cell is rather small (imagine a square of a road network in a big city.) Following the list solution, it is then noted that, at a given time instant t, the entire list of the cell has to be retrieved, which is large. In the primitive solution, instead, only the small list associated to the time instant t has to be retrieved.

On the other hand, however, this solution has an obvious drawback, the duplicate space consumption, due to the necessity of maintaining two structures.

A second, more crucial drawback, of this solution concerns Structure B, when a new time instant t is inserted into the upper index structure. Structure B implies that each time instant has its own list. Since we expect that during a short time period a small fraction only of the objects of a cell will move to another cell, we can easily create the list of the new time instant by copying and modifying slightly the list of the previous time instant. Obviously, when the lists are too long, we end up with a tremendous waste of space.

Thus, the primitive solution achieves the desired time complexity but it may also suffer from vast space consumption. Hence, it is necessary to maintain a large number of similar lists, in a space efficient manner. This is the reason why persistent indexing structures have been proposed.[8]

## 5. The advanced solution

There are certain application areas, which require storing and accessing all the versions in which a data structure has undergone. Such requirements have been identified in the seminal paper by Driscoll et al. [8], in which the notion of *persistent data structures* has been coined. More typically, consider a data structure D. If persistence is supported, all the versions $v_1, \ldots, v_{m+1}$ are maintained, as *D* undergoes a number of *m* update operations.

We identify two flavors of persistence, namely *partial* and *full persistence. I*n partial persistence, every version can be queried but only the most recent can be updated. In full persistence, every version can both be queried and updated. There is also a third kind, *confluent persistence.* Confluently persistent data structures support an operation, which combines two versions of the data structure to yield a new version.

Application of persistence to secondary memory data structures is of particular interest since persistence finds a fertile ground in databases. A simple example is that of transaction databases, which store data with a certain lifespan. An extensive treatment of temporal and bi-temporal DBs, as well as their relationship to persistence, can be found in the survey by Salzberg and Tsotras [14]. Lorentzos et al [9] have studied the creation and maintenance of versions at the database design level. The fully persistent case has been studied in [7]. In this paper, we make use of the partial persistent case. Two optimal, partially persistent B+ trees have already been developed, the Multi Version B-Tree (MVBT) by Becker et al. [1] and the Multi Version Access Structure (MVAS) by Varman and Verma [13]. Although they both share the same ideas, MVAS has a slightly better space consumption constant.

MVAS is a modified B+ tree. Its internal nodes contain index records and its leaves contain data records. A data record contains the fields [key, start, end, info], with their obvious meaning. An index record contains the fields [key, start, end, ptr], where ptr is a pointer to a node of the next level. The node pointed by the ptr pointer contains keys no less than *key*, has been created at the time instant *start* and has been copied at the time instant *end*.

A data record is *active* (live) if its *end* field has value '$', i.e. it has not been updated, deleted or copied to another node. If this is not the case, the data record is *inactive* (dead). An index record is *active* if it points to an sctive block at the immediately lower lever.

Figure 4 shows a possible instance of MVAS, and a simple scenario. At time 5 (upper part of the figure), the tree consists of two nodes, the root and one leaf, which contains all the data records. The figure shows that key A was inserted at time 1, key C was inserted at time 2 and was subsequently modified at times 3 and 4, and key B was inserted at time 5. Then, at time 6, key D is to be inserted. This insertion causes an overflow of the single leaf. Two new leaves are then created and the old leaf becomes inactive (all the records appear as shaded). The index record of the root, which points to the inactive leaf, also becomes inactive (shaded). The set of live records of the old leaf is sorted by key, is divided into two halves and each of these halves is copied to one of the two new leaves. Two new index records are created in the root. Their start value is the time at which the pointed leaves were created, i.e. time 6. Note that it is not always the case that two new leaves have to be created. For example, if instead of inserting key D, we had to update B at time 6, then the live records of the old leaf would fit in one new leaf.

To delete a record, we make use of a flag (shaded in Figure 4) and then count the remaining live records of the leaf. If they are too few, we may borrow some live records from a neighbor leaf, and create one or two new leaves.

We do not describe the operations of MVAS in further detail, because we want to give only the intuition behind this structure.

To search for a key x, at time t, we start from the root. We ignore records with a *start* value greater than t and an *end* value less than t. From the remaining records, we choose the one with the greatest key value, less than or equal to x. For example, if the search concerns key C, at the time instant 3, it is noted that only the inactive record (A, 1) at the root satisfies the time criterion, meaning that it was live at time instant 3. Following the pointer of this record, we reach the old inactive leaf, where we find that key C was really present at time 3.

The key idea is to maintain the following invariant: *For any version, the records contained in that version are sorted by key value and are clustered into secondary memory blocks in such a way, that each block contains B records belonging to that version.*

If *n* is the number of records in the current version and *k* is the output size, this invariant helps in achieving $O(\log_B n + k/B)$ I/Os (or block transfers) for search and update operations.

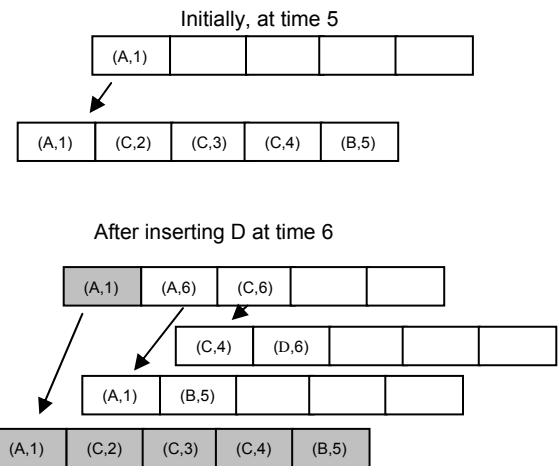

**Figure 4. A simple MVAS instance**

Based on the above, in our algorithm we use MVAS as the cell's indexing structure: Whenever an update occurs (one or more Ids either enter the cell or leave the cell), we perform the updates inside a cell, and create a new version of it. Versions are named by the time at which they are created.

Suppose now that we want to process the spatiotemporal predicate $(C_1, T_1)$. Assume also that, initially, $T_1$ is the time instant $t_1$. We search the indexing structure of cell $C_1$, in order to find all the leaves that were live at time $t_1$. Each such leaf contains from B/4 up to B records (this fact comes from the description of MVAS [13]), which belong to $F(C_1, T_1)$. According to the technical constraints of the structure, we charge each leaf with one additional I/O (We recursively traverse the tree in order to reach the desired leaves each of which is charged with the access of an internal node.) It follows that the total number of I/Os is not greater than $8*F(C_1, T_1)/B$ (for details see [13]). By using therefore a persistent indexing structure, we have managed to spare at most $O(F(C_1, T_1)/B)$ I/Os in order to retrieve $F(C_1, T_1)$.

Now assume that the time constraint $T_1$ of the spatio-temporal predicate is a time interval $[t_1, t_2]$. Then we can follow the *history* from time $t_1$ up to time $t_2$. When one or two leaves of the index structure die, either one or two new leaves are created. In either case, this death–birth sequence is triggered by update operations. When a leaf L dies we store into it a pointer to the newly born leaf. If two new leaves are created, we store into L two pointers. Figure 5 shows the leaves of the indexing structure at the time instants $t_i$, $t_{i+1}$, …, $t_{i+4}$.

As is shown in Figure 5, at time $t_i$, all the leaves have "experienced" insertions, as is shown by the shading). A series of deletions occur until time $t_{i+3}$. At this time, a number of insertions lead to a split of the remaining leaf. Suppose that we want the Ids of all the objects that entered the cell between the times $t_i$ and $t_{i+4}$. First, we retrieve the leaves that contain all the entries of time $t_i$. Then, by following the depicted pointers, we can retrieve all the desired leaves. In general, it is not definite that we will find new Ids for every leaf of the succeeding time instants. For example, in Figure 5, none of the leaves at time instants $t_{i+1}$ and $t_{i+2}$ contain new entries.

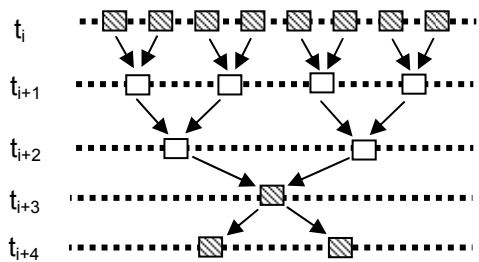

**Figure 5. Moving from $t_i$ to $t_{i+4}$**

Nevertheless, this fact does not cause any problem. The total number of accessed leaves is at most twice the number of the dead (shaded). Having in mind that each dead leaf contains at least B/4 Ids, which belong to $F(S_1, T_1)$, it follows that the total number of I/Os cannot be greater than $8*F(S_1, T_1)/B$. Having also in mind that we can reach the leaves of MVAS by applying a recursive procedure that accesses one internal node per leaf, we end up with a total number of $16*F(S_1, T_1)/B$ I/Os.

By use therefore of the advanced solution, we can process a spatiotemporal predicate $(S_1, T_1)$ by sparing at most $16*F(S_1, T_1)/B$ I/Os. The temporal constraint $T_1$ can be either a time instant or a time interval. We thus conclude that the query

$$Q_1 = \{(S_1,T_1),(S_2,T_2),\ldots,(S_r,T_N)\}$$

is proccessed in $O((max\{F(S_i, T_i)\}/B)*N)$ I/Os, meaning that, we have achieved our goal.

Beyond the theoretically excellent performance in terms of the number of block transfers (I/Os), this solution is also expected to achieve good results in terms of space consumption. Specifically, the indexing structure stores M versions, each of which is produced by one update, i.e. the space complexity is O(M).

## 5. Experimental Results

In this section we present the result of conducted experiments in order to compare the primitive and advanced solution with respect to the list solution. In particular, we have conducted an experimental study making the customary assumption that the disk page size is set to 512 bytes, the length of each key is 8 bytes, and the length of each pointer is 4 bytes. Consequently, each block contains B=42 elements. We use a relatively small page size so that the number of nodes in an index simulates a realistic situation with a large number of objects, A similar methodology has also been used in [15]. We generated synthetic data sets of moving object ids. The 2-dimensional spatial universe is a 1000x1000 grid, which simulates an actual universe of 1000 miles long in each direction. We also assume that we have a heavy traffic, generated by 1.000.000 vehicles. The velocity value distribution is skewed (zipf) towards 0 in range [0, 50]. The query cost is measured as the average number of node accesses in executing a workload of 200 queries with the same parameters. Implementations were carried out in the VC++ programming language.

The time efficiency of the primitive and the advanced solution, with respect to the list solution, is shown in Figure 6.

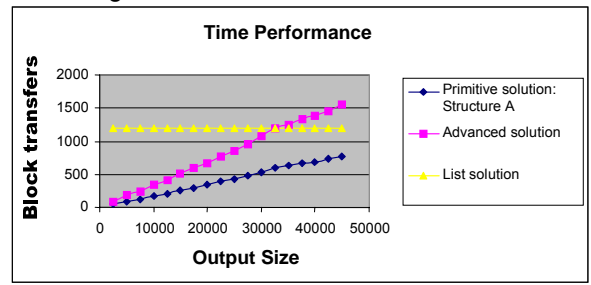

**Figure 6. Number of I/Os vs. Output size**

The average number of objects per cell was not more than 5000. The average number of predicates that appeared in the workload of the above queries was not more than 10. The output size of the queries varied in the range [2.500, 50.000].

The major advantage of the advanced solution, versus that of the list solution, becomes evident when the query output concerns only a small fraction of the contents of a cell. To make it clear, assume that a cell structure contains the data of the last week. Assume also that we seek all objects that were in the cell during a period of only a few hours. (Such cases occur close to the beginning of the x- axis of Figure 6.) As can then be seen in this figure, the number of I/Os is very low in the case of the advanced solution. As opposed to this, this number is very large in the case of the list solution, matching always the worst case, since the entire list must be extracted. Of course, if we want to extract all the objects that entered a cell during the last week, the advanced solution can be worse than the list solution. Note however that such queries are not realistic and are unlikely to be issued.

Comparing the number of I/Os between the primitive and the advanced solution, we conclude that they do not differ substantially. The primitive solution is better, requiring about half of the I/Os of the advanced. This is because the lists of structure A are optimally dense, i.e. all the disk pages that store a list are full, except for only the last page. On the other hand, the leaves of MVAS are not usually full. The penalty however for this superiority is the enormous space consumption of Structure B (see Section 4).

The theoretical space complexity of MVAS is $O(N/B)$ blocks, where N is the number of stored elements and B the block size. In Figure 7 we have plotted the consumed space of (a) only structure A of the primitive solution (recall that the space of structure B makes the primitive solution unpractical and far worse than the other two solutions), (b) the advanced solution and (c) the list solution. As can be proved, the three competitors have theoretically identical space complexity. In Figure 7, it can be seen that the precise consumed space of the three competitors differs by a constant multiplication factor.

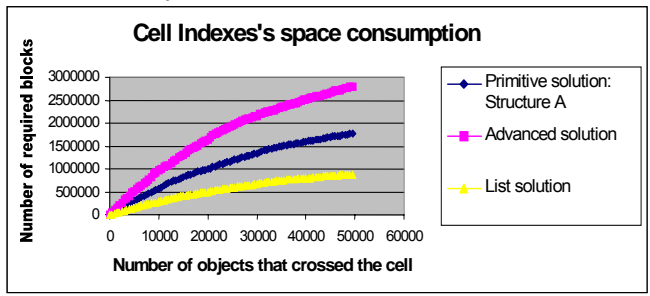

**Figure 7. Space consumption of cell's indexing schemes**

From Figure 7 it also follows that Structure A consumes less space than the advanced solution, but not less that half of it. The increased space consumption of MVAS has been expected, because it is a complicated index structure, requiring more pointers and other pieces of data compared to those of the B-tree of Structure A. Finally, the list solution is the best, and this was expected since lists occupy less space than trees. Recall, however, that the increased space consumption of the advanced solution leads to a dramatically improved time efficiency on realistic queries, as has been shown in Figure 6.

## 6. Discussion

We have dealt with the problem of efficient processing of spatio-temporal predicates. For our purposes, a previous approach [6], which uses a grid, and associates a list with each cell of the grid, was extended by integrating into it a persistent indexing structure. This indexing structure enables the efficient maintenance of versions of the lists that are created by update operations. Since the versions correspond to time instants, what we have finally achieved is to *hide* the time dimension into a persistent indexing structure. While achieving time efficiency, we have thus managed to maintain the same space complexity with that in [6].